\documentclass[runningheads]{llncs}
\usepackage[T1]{fontenc}
\usepackage{graphicx}
\usepackage{booktabs} 
\usepackage{longtable}
\usepackage{tabularx}
\usepackage{xcolor}
\usepackage{amsmath}
\usepackage[numbers,sort&compress]{natbib}
\usepackage{xspace}
\usepackage[capitalize]{cleveref}
\usepackage{todonotes}
\usepackage{siunitx}

\newcommand{\dgr}[1]{\SI{#1}{\degree}}

\makeatletter
\newcommand{\printfnsymbol}[1]{%
  \textsuperscript{\@fnsymbol{#1}}%
}
\makeatother

\definecolor{frostfairy}{HTML}{B7CAF5}
\definecolor{burningorange}{HTML}{FA6C25}
\definecolor{sizzlingred}{HTML}{FA3C56}
\definecolor{splashofgrenadine}{HTML}{FA80DE}
\definecolor{goldenglitter}{HTML}{F9E574}
\definecolor{retroblue}{HTML}{2D5FFA}
\definecolor{phosphorescentblue}{HTML}{1CF0EB}
\definecolor{banafšviolet}{HTML}{621795}
\definecolor{blankagreen}{HTML}{9FD63F}
\definecolor{forcefulorange}{HTML}{F5950F}
\definecolor{felt}{HTML}{277442}

\newcommand{\codename}{AI-Spectra\xspace}

\begin{document}
%
%\title{Model Multiplicity in Interactive Systems}
\title{\codename: A Visual Dashboard for Model Multiplicity to Enhance Informed and Transparent Decision-Making}
\titlerunning{\codename: A Visual Dashboard for Model Multiplicity}

%
%  https://chatgpt.com/share/66ea93e7-185c-800d-b0a4-ca89659a77b0
%
%
%
%\titlerunning{Abbreviated paper title}
% If the paper title is too long for the running head, you can set
% an abbreviated paper title here
%
\author{
Gilles Eerlings\orcidID{0009-0005-7765-6774}\thanks{equal contribution} \and
Sebe Vanbrabant\orcidID{0009-0001-7996-6048}\printfnsymbol{1} \and 
Jori Liesenborgs\orcidID{0000-0002-3648-8031} \and 
Gustavo Rovelo Ruiz\orcidID{0000-0001-7580-8950} \and 
Davy Vanacken\orcidID{0000-0001-8436-5119} \and
Kris Luyten\orcidID{0000-0002-4194-1101} 
}
\authorrunning{Eerlings G., Vanbrabant S., et al.}
% First names are abbreviated in the running head.
% If there are more than two authors, 'et al.' is used.
%
%\institute{Hasselt University - Flanders Make, Expertise Centre for Digital Media\\
\institute{Hasselt University - Flanders Make, Digital Future Lab
\email{firstname.lastname@uhasselt.be}}
\maketitle              % typeset the header of the contribution
%

% \begin{abstract}
%     The proliferation of artificial intelligence (AI) in interactive systems has led to significant challenges in model integration, but also end-user-related aspects such as over- and undertrust. This paper presents an initial exploration on how multiple AI models with the same performance and behavior but different internal workings--a phenomenon called model multiplicity--affect system integration and user interaction. We discuss the implications of model multiplicity for transparency, trust, and operational effectiveness in interactive software systems.
% \end{abstract}

\begin{abstract}
%previous version:
    %Model multiplicity—where multiple AI models achieve similar performance through different internal mechanisms—presents unique challenges and opportunities for interactive systems. This paper explores a novel ``many simultaneous expert advisors'' approach to leverage model multiplicity, moving beyond reliance on single models to enhance system robustness and user trust. We present an initial framework for training and deploying multiplicitous models in interactive systems, addressing key challenges in visualization and user interaction. Using the MNIST dataset as a case study, we demonstrate how various model configurations can be effectively represented and compared using a custom adaptation of Chernoff faces, which we call ``Chernoff Bots''. This visualization technique lets users quickly interpret complex, multivariate model configurations and compare predictions across multiple models. We analyze the implications of our approach through the lens of established Human-AI Interaction guidelines. We identify promising directions and challenges for future work, particularly in scaling model multiplicity to more complex systems while maintaining interpretability and computational efficiency. Our work contributes to the growing discourse on making AI systems more transparent, trustworthy, and effective through the strategic use of multiple models.

%handwritten long version:
     We present an approach, \codename, to leverage model multiplicity for interactive systems. Model multiplicity means using slightly different AI models yielding equally valid outcomes or predictions for the same task, thus relying on many simultaneous ``expert advisors'' that can have different opinions. Dealing with multiple AI models that generate potentially divergent results for the same task is challenging for users to deal with. It helps users understand and identify AI models are not always correct and might differ, but it can also result in an information overload when being confronted with multiple results instead of one. \codename leverages model multiplicity by using a visual dashboard designed for conveying what AI models generate which results while minimizing the cognitive effort to detect consensus among models and what type of models might have different opinions. We use a custom adaptation of Chernoff faces for \codename; Chernoff Bots. This visualization technique lets users quickly interpret complex, multivariate model configurations and compare predictions across multiple models. Our design is informed by building on established Human-AI Interaction guidelines and well know practices in information visualization. We validated our approach through a series of experiments training a wide variation of models with the MNIST dataset to perform number recognition. Our work contributes to the growing discourse on making AI systems more transparent, trustworthy, and effective through the strategic use of multiple models.

%Shorter version, by GPT-4o:
     %We present \codename, an approach that leverages model multiplicity—slightly different AI models yielding equally valid outcomes—for interactive systems. While model multiplicity enhances users’ understanding of AI variability, it can overwhelm them with divergent results. \codename addresses this by providing a visual dashboard that reduces cognitive load, allowing users to easily detect consensus and outliers among models. We introduce Chernoff Bots, an adaptation of Chernoff faces, to visually represent multivariate model configurations and facilitate comparison across models. Grounded in Human-AI Interaction guidelines and information visualization practices, \codename is validated through experiments on the MNIST dataset. Our work advances AI transparency, trustworthiness, and effectiveness by strategically integrating multiple models.
\end{abstract}

%Generative AI can generate various types of data (images, text, layout, persona's,\ldots) driven by models that are trained with specific training data. Generative AI is best known for its capability to generate text and images, but is also becoming more common as a supporting tool during the engineering life cycle for building interactive systems. Researchers have started exploring the use of Large Language Models for various aspects of the Human Centred Development Process \cite{HCIGPT24}. It has been used, a.o., to do data-driven designs and generating prototypes \cite{webzeitgeist13,Rico17,bayesiangrammer12}, integrating multi-modal interaction techniques \cite{yang2024reactgenie} and even feedback on User Interface designs \cite{LLMFeedback23} and usage within a broader social context \cite{simulcra22}.

% \section{Visual Dashboard + Links}
% \url{https://chatgpt.com/share/66ea93e7-185c-800d-b0a4-ca89659a77b0}

% \url{https://www.kaggle.com/code/delai50/visualize-numerical-features-with-chernoff-faces}

\section{Introduction}
% Probleemstelling
Integrating artificial intelligence (AI) models into interactive software systems is becoming increasingly common across sectors such as healthcare and customer service. Typically, a single AI model is embedded within an interactive system, requiring careful design to ensure accessibility and usability for end users. Relying on a single model might quickly lead to a miscalibration in trust, as there is a limited basis for comparison and often a low degree of transparency due to the lack of explanations. This lack of trust and transparency can be problematic, as both are crucial aspects of effective human-AI collaboration~\cite{asan2020artificial,coppers_intellingo_2018,coppers-fortniot,explainer}.

% Many experts als mogelijke oplossing?
To alleviate these issues relating to trust and transparency, we use a `\emph{many simultaneous expert advisors}' approach in which multiple AI models, each considered an expert advisor, are used simultaneously within an interactive system to offer users more balanced and well-substantiated advice. %This might involve using multiple black-box and white-box models together or employing similar models (e.g., in terms of architecture and interpretability) with varying hyperparameter configurations. %In our work, we focus on the process of preparing and using closely related and very similar models for a single task and making this complex information accessible for end-users through a visual dashboard for decision support. %Unlike traditional model multiplicity, which often focuses on binary classification, our work involves n classes, extending the application of model multiplicity to more complex, multi-class scenarios.
There are often multiple variations of AI models trained for the same task that achieve similar performance during the validation. This phenomenon, also known as \emph{model multiplicity} (MM)~\cite{ModelMultiplicity}, can be of great value to correct \emph{over- and undertrust} in an AI system. Having a limited understanding of AI can be frustrating, causing users to lose trust~\cite{antifakos_towards_2005,muir_trust_1994}. More trust, however, is not always better, since users might trust the system even when it is not behaving as intended~\cite{yang_how_2020}. We aim to move away from a single AI model, thus one single output, and evolve toward multiple AI models that provide multiple possible outputs and the context on which their output is based.

% Elaborate on many experts
The \emph{many simultaneous expert advisors}, or model multiplicity, approach is particularly valuable for critical decision support systems, such as those used in medical diagnostics, financial forecasting, and urban planning. In the healthcare sector, for instance, employing multiple AI models allows for a more comprehensive analysis of patient data by cross-verifying diagnoses, thereby reducing the risk of error. This not only mitigates the risks associated with over-reliance on a single model but also provides a platform for continuous learning and improvement of AI systems. Rather than focusing on making the individual models transparent and interpretable~\cite{rudin2019stop, zachary-mythos2018}, we increase trust and transparency by informing users about the differences between multiple expert advisors. Expert advisors are often black box models that are known to perform well for specific complex problems, but they are very hard to nearly impossible to explain to end users, however. Analyzing the differences between multiple similar models and their respective outputs can potentially lead to the refinement of the algorithms that are used, improvement of accuracy and reliability, and the uncovering of problematic patterns in the training dataset. This approach enhances transparency and builds trust among users by ensuring that decisions are thoroughly considered and vetted through multiple expert advisors (i.e., the various models).

% Paper structure, idea, goal and outcome(s)
%This paper explores the `many experts' through the lens of the `model multiplicity' phenomenon. 
%Our goal is to provide deeper insights into the behavior of these models and their predictions. To achieve this, 
We introduce \codename; a process and visual dashboard to leverage model multiplicity to enhance informed and transparent decision-making. We focus on classifier models, often driven by neural networks and deep learning. This is particularly interesting since these models are considered black box models that behave in ways that are hard or even impossible to explain. Our contribution is threefold:
\begin{itemize}
    \item[(1)] We define \emph{a process for preparing, training and identifying suitable sets of AI models} for effective model multiplicity.
    \item[(2)] We design \emph{a compact and easy-to-recognize visual representation that conveys the hyperparameters, properties and background of an AI model} to the user. For this purpose, we leverage the Chernoff faces technique \cite{chernoff1973novel, flury81-chernoff} and tailor it to AI models.
    \item[(3)] We present a \emph{visual dashboard that can present the distribution of answers across multiple models} in relationship to the varying backgrounds and properties of these models, like the overview depicted in \cref{fig:highly_confident_rashomon_set}. 
\end{itemize}
We validate \codename with classifier models that are trained on the MNIST dataset, consisting of handwritten digits for training for number recognition. The visual dashboard of \codename is depicted in  \cref{fig:highly_confident_rashomon_set} as a set of Chernoff faces presented as a bar chart. Each bar represent a specific prediction for the MNIST dataset (10 possible digits that can be recognized), and each Chernoff bot in the bar represents an AI model that supports that specific prediction. This makes it easy to compare the number of models that support a prediction, as well as to recognize to what predictions both outlier as well as comparable models contribute.
% \begin{figure}[htb]
%     \centering
%     \includegraphics[width=\textwidth]{images/highly_confident_rashomon_set.png}
%     \caption{The visual dashboard of \codename is depicted as a set of Chernoff faces presented as a bar chart. Each bar represent a specific prediction, and each Chernoff bot in the bar represents an AI model that supports that specific prediction. This makes it easy to compare the number of models that support a prediction, as well as to recognize to what predictions both outlier as well as comparable models contribute.}
%     \label{fig:AISpectraBarChart}
% \end{figure}

%Following this, we present the implementation of our interactive system, which compares multiple models using Chernoff faces. Finally, we highlight the opportunities and challenges associated with integrating multiple AI models into a single system.

\section{Related Work}

\subsection{Model Multiplicity}\label{sec:modmul}

When training an AI model for a task, multiple decisions are made about how the model should be constructed (e.g., hyperparameters, training data), resulting in several possible variations of models trained for the same purpose. Traditionally, the model that achieves the highest accuracy for that task is chosen. However, this process falls short when multiple models trained for the same task achieve similar accuracies in different ways due to their distinct decision boundaries ~\cite{Ganesh2023AnEI}, which are the surfaces formed by a model's learning process to separate classes within the feature space. This phenomenon is known as \emph{model multiplicity} (MM) or \emph{the Rashomon effect}~\cite{Breiman2001}, and the set of models that conform to model multiplicity for a certain task is referred to as the \textit{Rashomon set} having \textit{multiplicitous models}~\cite{fisher2018}. Model multiplicity was previously leveraged to prioritize other desirable properties beyond accuracy during the model selection process, such as interpretability or fairness, by choosing the model from the Rashomon set that prioritized this desired property~\cite{ModelMultiplicity}. While this prioritization focuses on the individual strengths of these equivalent models, equivalent models may make different choices, potentially confusing users. This can lead to issues of both \emph{over- and undertrust} in the AI system. Users may lose trust when they cannot understand why one model's predictions are favored over another ~\cite{antifakos_towards_2005,muir_trust_1994}, or conversely, they may place too much trust in the system, even when it is not performing as expected ~\cite{yang_how_2020}. Both scenarios can negatively impact the user’s interaction with the system.

Related work has also addressed procedural multiplicity, which occurs when models within a Rashomon set produce consistent predictions for a given input despite having different internal logic~\cite{brunet2022implications}. Because of these internal differences, local explanations such as those generated by SHAP~\cite{lundberg2017unified} or LIME~\cite{ribeiro2016why} will also vary between models, even in the case of procedural multiplicity, a situation referred to as \emph{explanatory multiplicity}. As these explanations could be inconsistent and even contradicting, the same problems related to trust will also occur. Therefore, there is a need for a global explanation framework that can effectively communicate the differences between the models in a Rashomon set.

% Nu over dat die aanpak van flexibel modellen kiezen eigenlijk niet veel verandert aan appropriate trust en dan onze approach dat we alle sterktes van de verschillende modellen willen combineren
We aim to move beyond relying on a single AI model and its singular output, instead advancing toward a framework where multiple AI models provide diverse outputs along with the context behind each prediction. Integrating model multiplicity into interactive systems enhances reliability, enables more nuanced interactions with AI, and introduces significant complexity in system engineering. Model multiplicity demands careful orchestration to ensure that the outputs from various models are integrated coherently, transparently, and in a user-friendly manner. Engineers must design systems capable of handling potentially conflicting advice from different models and determine how to prioritize or merge these outputs effectively. This involves creating advanced decision-making algorithms or voting systems that transparently assess and integrate diverse outputs. At the same time, the user interface must communicate the rationale behind multiple AI-driven decisions clearly and intuitively.

\subsection{Visualizing and Comparing Multiplicitous Models}
% TimberTrek
Exploring the outputs of multiple AI models presents a challenge. Many tools have been developed to help users assess machine learning models through interactive visualizations~\cite{explainer,wang2023gamcoach,wexler2020whatif}. However, one lesser-researched area is \textit{model comparison}, as opposed to model interpretation. Most machine learning tools support comparing different models' parameterizations by summarizing performance statistics, but they do not show the models themselves as they evolve during (interactive) training~\cite{amershi2010examining,gleicher2020boxer}. Only a few visualization tools, such as TimberTrek~\cite{TimberTrek}, facilitate the exploratory comparison of multiple models in a way that is accessible to end users. These tools provide intuitive dashboards for exploring multiple model results, and their rich visualizations are primarily used as stand-alone systems designed for deep exploration through visual explanations. However, TimberTrek is limited to decision trees, making it unsuitable for systems that do not use this specific model architecture.

% CueFlik, een systeem dat toelaat interactief modellen te trainen met een history en de impact van de changes laat zien om een model te verbeteren
% Concrete delta die we kunnen geven is dat het bij CueFlik zinvol is dat die history bestaat zodat mensen hun huidige model kunnen vergelijken met een historisch model en zo betere resultaten krijgen door dat vergelijken tussen modellen
An early approach to comparing multiple models was implemented using CueFlik, a system that allows end users to train visual concepts for re-ranking web image search results based on their visual characteristics~\cite{amershi2010examining,fogarty2008cueflik}. In this system, users are provided with a historical overview of how different labels for objects influence the system's behavior in relation to their goals. By incorporating history and revision mechanisms, CueFlik enabled users to achieve better final models within the same timeframe. These mechanisms allowed users to compare the current model with any previous version and backtrack if the model appeared to deviate from the desired outcome.

% Model comparison gebaseerd op modellen getraind op verschillende subsets van data
Aggregated measures provide a quick summary of model performance but often lack the detail needed to examine performance on different subsets of data. To address this limitation, the Boxer system was developed~\cite{gleicher2020boxer}. Boxer allows for a detailed examination and comparison of decision tree classification models trained on different subsets (i.e., boxes) of a dataset. Users can interactively compare selected models on various aspects, such as model performance and data distribution, using various visualizations, including bar charts, confusion matrices and histograms.

% Generalization of other methods
Other approaches present multiple models in more traditional ways, such as graph plots and bar charts that display model metrics side by side~\cite{vandenelzen2011baobabview,visual-multiple-models,muhlbacher2018treepod,Murugesan2019DeepCompare,ElAssady2018Progressive}. These systems are typically designed to focus on model selection, aiming to optimize performance or identify the most suitable model for a specific task. Furthermore, many of these systems rely on tree-based AI methods, which are generally easier to interpret but often less accurate than black-box models~\cite{caruana2015intelligible}. Therefore, further research is needed to address model multiplicity for black-box models effectively, focusing on model comparison rather than model selection. Furthermore, our approach to model multiplicity emphasizes the concurrent use of all models and their predictions, as we want to rely on a council of equally accurate experts. This contrasts with approaches that consult a specific model or aggregate outputs, such as those discussed by Black et al.~\cite{ModelMultiplicity}. Concretely, we envision an approach that enables users to visually assess this council of multiplicitous models and their predictions so that they can form their own opinions on the predictions and the models themselves.

\section{Training, Preparing and Executing Model Multiplicity}
\begin{figure}[htb]
    \centering
    \includegraphics[width=\textwidth]{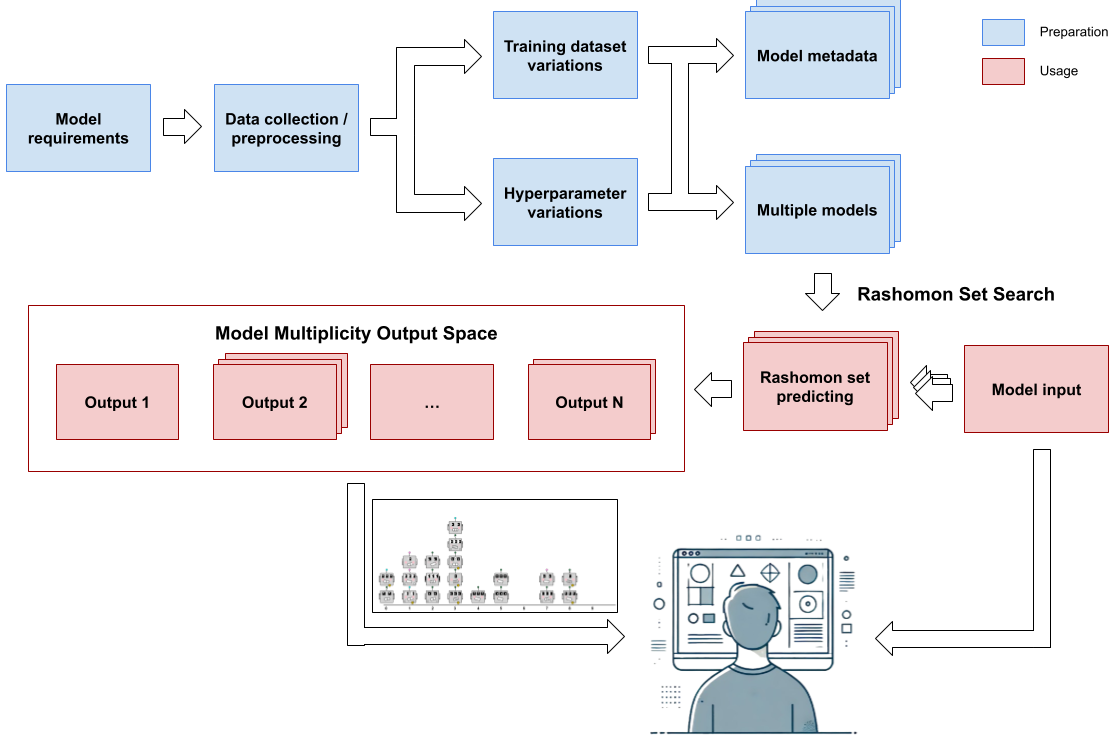}
    \caption{A process for using and integrating model multiplicity, highlighting two stages: training and preparation (blue) and usage (red). This process outlines the steps from model requirements and data preprocessing through variations in training datasets and hyperparameters, ultimately resulting in multiple models. From these models, Rashomon sets are identified, allowing comparison and selection for predictions. While not directly integrated into the pipeline, the model metadata links and keeps track of the various configurations of their respective models.}
    \label{fig:eis-for-mm}
\end{figure}

Building upon the frameworks provided by Amershi et al.~in Software Engineering for Machine Learning~\cite{SEforML}, we designed an initial process for \codename that integrates these principles with a model multiplicity approach. This process is illustrated in \cref{fig:eis-for-mm}. It consists of two stages for implementing model multiplicity: the preparation stage and the usage stage.
%For the `usage' stage, new ways to deal with multiple outcomes need to be designed for end-users to be able to process the multiple outcomes origination from the AI subsystem.% In the future, we will explore explainable AI components to improve appropriate trust and understanding of model multiplicity outcomes (e.g., Spinner et al.~\cite{explainer}).

% Eventueel ook diezelfde terminologie uit de figuur (preparation vs usage) laten terugkomen

\paragraph{\textbf{Model Preparation}}
The preparation stage begins with defining the model requirements and aligning the problem at hand with the capabilities of the machine learning model by determining which models are most suitable for the given task. Afterward, data must be collected, cleaned, and labeled to ensure usability. If necessary, feature engineering is applied to improve data quality. Following these steps, the `many expert' candidate models are trained. To distinguish between models trained for the same purpose, we consider variations in 1) the input data for the models, 2) the architecture of the models, and 3) the training process. Specifically, models are generated by applying different configurations to the same dataset, trained with varying parameter settings (such as the number of layers and neurons per layer in a neural network), and the training processes are modified (e.g., varying the number of epochs, batch sizes, or optimizers). This approach resembles a grid search; however, instead of seeking a single optimal model, we aim to retain all models with their respective configurations, stored as the `model metadata'. This metadata is used to keep track of which model has which configurations applied to them. Next, we search for Rashomon sets with the goal of grouping models that correctly predict the label for a given set of samples, thereby creating groups of models that `think' alike. To form these groups, each sample from the test set is evaluated by all models sequentially. Based on the predictions and associated accuracies, a filtering process is applied to eliminate underperforming models or `bad learners'. Models that achieve high accuracy are retained and grouped together, forming a Rashomon set.

\paragraph{\textbf{Model Usage}}
Running multiple models simultaneously produces a range of outcomes that need to be aggregated. Since we prioritize insights from `many experts' rather than relying on a single prediction, it is essential to present this variety of predictions without reducing them to a single value, such as an average. However, obtaining a \textit{visual} overview of each model's prediction for different classes can be challenging, especially as the number of possible
outcomes grows. For n-class classification models like neural networks, where outputs span numerous possible classes, the visualization will be more complex than visualizing the solution space of models performing binary classification. That said, explaining how each model arrives at its specific conclusion can complicate visualizations even further, especially depending on the type of model used (e.g., neural networks versus decision trees), since the decision process would also have to be integrated.

\section{Engineering A Visual Dashboard for Comparing Multiplicitous Model Sets}
\label{sec:dashboard}

\subsection{Challenges in Visualizing Model Multiplicity in Interactive Systems}
When integrating multiple model outputs into a user interface, presenting these outcomes in a meaningful and interpretable way poses significant challenges, particularly for complex models like neural networks. This complexity arises because different models may produce varying results for the same input, making it difficult for users to understand the differences or trust the predictions. To address these issues, we turn to the Guidelines for Human-AI Interaction~\cite{guidelines-humanai} as a reference framework. These guidelines provide valuable insights into designing interfaces that support effective human-AI collaboration, ensuring that users can interpret, trust, and act on the aggregated predictions from multiple models. 

\subsection{Applying Human-AI Interaction Guidelines to Model Multiplicity}
We selected six guidelines on which model multiplicity has the most significant impact with respect to when only a single AI model is used:
\newcounter{nrguideline}
\setcounter{nrguideline}{0} 
\newenvironment{guideline}[1][]{\refstepcounter{nrguideline}\par\smallskip
   \noindent\textbf{\thenrguideline. \textbf{#1}} \rmfamily}{\smallskip}
\begin{guideline}[(G3) Time services based on context.]
Model multiplicity introduces additional computational demands, which can lead to delays in system responsiveness. An interactive system must strategically manage when and how to engage users to prevent interruptions at inopportune moments. By optimizing the timing of prediction requests and displays according to the user's current task, the system can effectively \textbf{balance the computational load of multiple models while preserving a positive user experience}.
\end{guideline}
\begin{guideline}[(G6) Mitigate social biases.]
Model multiplicity aims to deliver more balanced results and assist in detecting bias across different models. An interactive system that incorporates model multiplicity should \textbf{facilitate the identification of biased models}. By providing visual comparisons of model outcomes, the system can help users recognize biases more effectively, ultimately guiding better decision-making.
\end{guideline}
\begin{guideline}[(G11) Make clear why the system did what it did.] 
Explaining the behavior of a system using multiple models can be challenging, especially when employing black-box models such as deep neural networks. The interface should enable users to explore the reasoning behind each model's output to enhance transparency. Interactive dashboards that allow users to \textbf{investigate the specific outcomes produced by various models} can facilitate a better understanding of the reasoning behind each model's decisions.
\end{guideline}
\begin{guideline}[(G14) Update and adapt cautiously.]
Updating model multiplicitous systems can be challenging, as each model may respond differently to changes. Such updates might introduce inconsistencies or affect performance in ways that eliminate model multiplicity when the accuracy of models becomes drastically different, unlike in single-model systems. Therefore, an interactive system should \textbf{enable users to compare similar models that behave differently} to assess the potential impact of updates on overall performance.
\end{guideline}
\begin{guideline}[(G15) Encourage granular feedback.] 
Collecting feedback in a model multiplicity system is more complex, as different models may respond differently to the same input. The system should \textbf{facilitate users in reflecting on individual model outputs} rather than providing generalized feedback. This approach enables more targeted improvements, allowing for the refinement of model selection based on user observations.
\end{guideline}
\begin{guideline}[(G16) Convey the consequences of user actions.] 
In systems with multiple models, user actions can impact each model differently, making it a complex and time-consuming process. As a result, immediately conveying the consequences of user actions may not be feasible. Instead, the system should \textbf{give users insights into how their actions influence each model over time}, allowing for a more informed decision-making process.
\end{guideline}
\medskip

Our focus is on leveraging model multiplicity to enhance informed and transparent decision-making, for which guidelines \emph{(G6) ``facilitate the identification of biased models''}, \emph{(G11) ``investigate the specific outcomes produced by various models''}, \emph{(G14) ``enable users to compare similar models that behave differently''} and \emph{(G15) ``facilitate users in reflecting on individual model outputs''} are particularly important. Guideline \emph{(G3) ``balance the computational load of multiple models while preserving a positive user experience''} is an important issue since the computational demand increases significantly, however out of scope for this work. Guideline \emph{(G16) ``give users insights into how their actions influence each model over time''} is relevant when using AI that has embedded interaction with the user, or uses the context of use. However this is not the case for the classifier models we work with.

% Hier komt het gedeelte over de Chernoff faces: design voor comparison en ook de betekenissen die we hebben gegeven aan de verschillende aspecten van de robot

\subsection{Chernoff Bots for Multidimensional Data Representation}
% Design van de faces + korte introductie over Chernoff faces
In this section we introduce a technique to visualize the variations of multiple models at once, enabling quick comparisons between models and their behavior. We use Chernoff faces~\cite{chernoff1973novel, flury81-chernoff} to graphically represent points in k-dimensional space. Chernoff faces leverage our innate ability to recognize differences in human-like faces, even when these are rendered in a cartoon style. They serve as mnemonic devices to reveal complex relationships that are not immediately apparent. Moreover, Chernoff faces \textbf{guarantee the presence of positive hedonic aspects} for human users by leveraging this cognitive familiarity of facial recognition, fostering intuitive pattern recognition, memorability, and enjoyable, exploratory interaction with data. This approach offers a powerful tool for enhancing human-in-the-loop systems by using facial features to represent multidimensional data. This approach takes advantage of human intuition and our natural ability to recognize facial patterns, making it particularly effective for identifying differences in complex models, such as machine learning models. By mapping various model parameters to distinct facial characteristics, users can quickly and easily spot variations between models, facilitating faster and more intuitive comparisons. 

%Chernoff faces can enhance human-in-the-loop systems by leveraging human intuition and flexibility to visually interpret complex data patterns, making it easier to differentiate between multidimensional data represented through facial features.

\begin{figure}[htb]
    \centering
    \includegraphics[width=\textwidth]{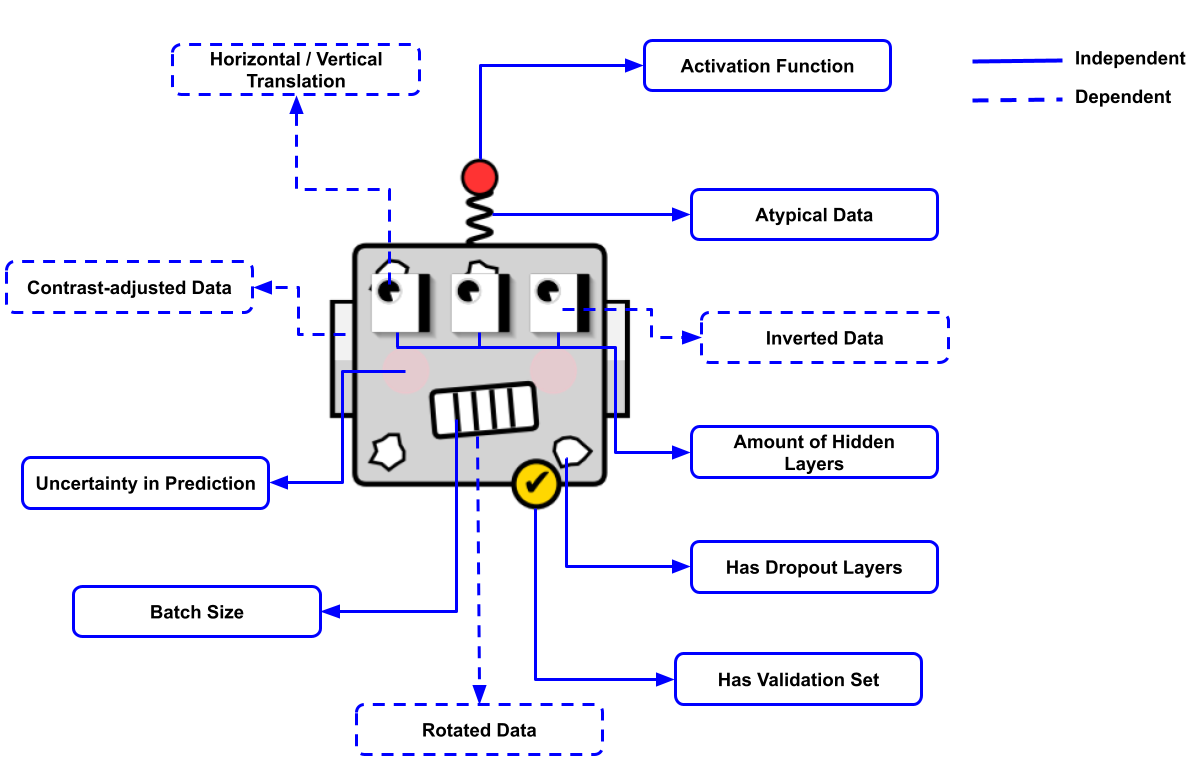}
    \caption{Example of a Chernoff Bot, featuring callouts describing various design aspects. Each variation referenced in \cref{tab:model_variations} has been thoughtfully translated into distinct features of the robot, illustrating how different data attributes are represented visually.}
    \label{fig:chernoffbots_linked}
\end{figure}

% Link met de verschillende parameters van het model en de training
%Although the primary goal of using Chernoff faces is to quickly represent multidimensional data at a glance, meaning that specific human aspects do not need to be linked to aspects of the data, we opted to assign meaningful interpretations to each variation to facilitate identification on top of comparison. To this end, 
We developed a custom approach based on Chernoff faces: our visual representation is tailored to the type of (meta-)data of machine learning models. We use depictions of robot faces, dubbed a ``Chernoff Bot'', instead of a human face to allow for greater creative freedom in introducing unconventional features, such as varying the number of eyes. This choice lets us explore designs that do not need to adhere to realistic human proportions. We aimed to map variations in the data directly to various parts of the robot's face wherever applicable. The Chernoff Bots capture two variations of the preparation process outlined in \cref{fig:eis-for-mm}: variations in training data and differing model configurations. Since the dataset is unique to each problem to be solved, we categorize the aspects into independent and dependent factors. Independent factors are not specific to any particular problem and are instead tied to the model architecture, in this case, neural networks. Dependent factors, on the other hand, are data-specific factors and cannot always be transferred to other problems. These dependent factors allow model developers to create custom representations of problem-specific data within a Chernoff Bot. We will discuss how we transferred our MNIST dataset-dependent variations into Chernoff representations in \cref{sec:usecase:preparation}. A complete overview of these mappings is presented in \cref{tab:variations_chernoff}, along with an example of a face shown in \cref{fig:chernoffbots_linked}. We decided to omit the optimizer from this representation, as its impact on the classifier was negligible.
\begin{table}[ht]
    \centering
    \caption{Overview of the different variations on the training data, model parameters, and training process and their respective representation in the robot's face.}
    \label{tab:variations_chernoff}
    \begin{tabularx}{\textwidth}{lX}
        % \hline
        
        \textbf{Variation}          & \textbf{Representation in Robot} \\
        \hline
        \multicolumn{2}{c}{\textit{Dataset Independent}} \\
        \hline
        Outlier Percentage          & Curvature of the antenna \\
        Typical Percentage          & Straightness of the antenna \\
        Number of Hidden Layers     & Number of eyes \\
        Dropout                     & Presence of holes in the body \\
        Activation Functions        & Color of the antenna: \textcolor{frostfairy}{Frost Fairy}, \textcolor{burningorange}{Burning Orange}, \textcolor{sizzlingred}{Sizzling Red}, \textcolor{splashofgrenadine}{Splash of Grenadine}, \textcolor{goldenglitter}{Golden Glitter}, \textcolor{retroblue}{Retro Blue}, \textcolor{phosphorescentblue}{Phosphorescent Blue}, \textcolor{banafšviolet}{Banafš Violet}, \textcolor{blankagreen}{Blanka Green}, \textcolor{forcefulorange}{Forceful Orange} or \textcolor{felt}{Felt}. \\
        Batch Size                  & 0, 1, 2, 3 or 4 lines in teeth (i.e. $ \log_2(\text{batchSize}) - 5 $) \\
        Optimizer                   & / \\
        Validation Set Usage        & Presence of checkmark badge \\
        \hline

        \hline
        \multicolumn{2}{c}{\textit{Dataset Dependent}} \\
        \hline
         Horizontal Translation      & Horizontal displacement of the pupils \\
        Vertical Translation        & Vertical displacement of the pupils \\
        Rotation                    & Rotation of the mouth \\
        Contrast Manipulation       & Contrast-adjustment of the ear-piece \\ 
        Image Color Inversion       & Color inversion of the eye and pupil \\
        % \hline
    \end{tabularx}
\end{table}

% Weergeven van de faces in een mixed-initiative bubble chart scatter plot, aanhalen hoe we het zien namelijk als een outlier detection en door het in bar chart formaat te steken halen we in 1 opzicht het aantal faces op en zien we meteen welke labels door welke modellen worden gepredict,

\subsection{Integrating Chernoff Bots into an Interactive Dashboard}
\begin{figure}[!h]
    \centering
    \includegraphics[width=\textwidth]{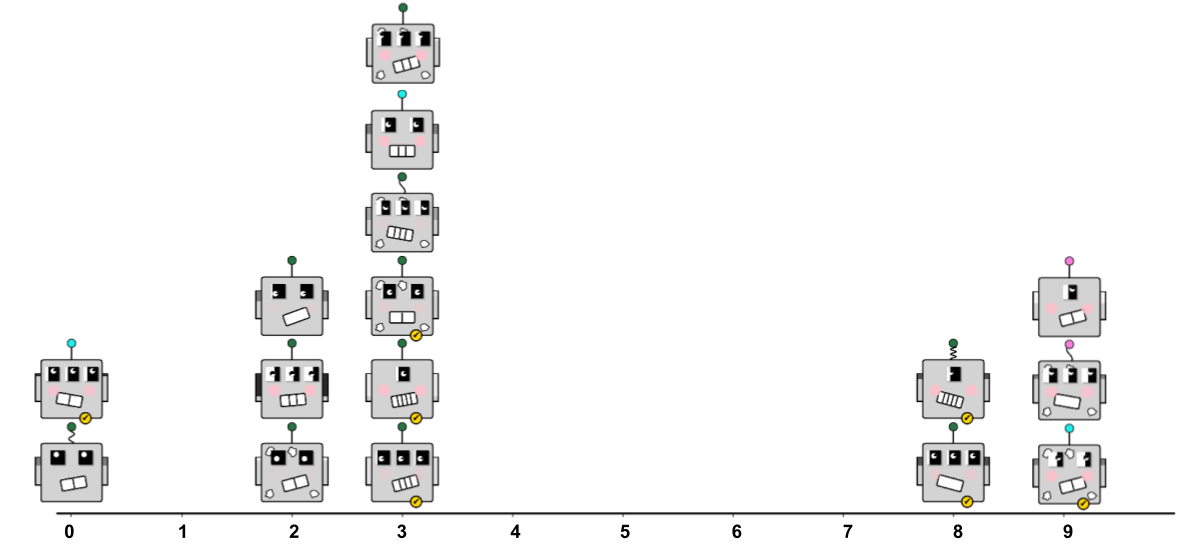}
    \caption{An example bar chart using Chernoff Bots to represent each predicted sample. The x-axis shows the possible labels for the task. The height of each bar indicates the frequency at which each label was predicted so that users can get an immediate impression of the most agreed-upon outputs.}
    \label{fig:highly_confident_rashomon_set}
\end{figure}

As we work with a model multiplicity system, we integrate multiple Chernoff Bots into a visualization so that users can explore the model multiplicity output space (depicted in \cref{fig:eis-for-mm}). Concretely, we create a bar chart out of Chernoff Bots for each predicted sample as shown in \cref{fig:highly_confident_rashomon_set}. The x-axis of this bar chart displays the possible labels for the task (in the case of the MNIST dataset, the ten possible digits). Whenever a model predicts a particular label, its Chernoff representation is stacked on that label, forming a bar in the y-axis. The height of the resulting bar corresponds to the number of models that predicted the same label for that input sample. This grouping of similar `thinking' experts allows the user to get an immediate impression of what output is most agreed upon and, therefore, assumed most likely to be correct.
Furthermore, users can form their own judgments by observing the input data and comparing the differences between the Chernoff Bots.

\section{Application to the MNIST Dataset}
\label{sec:application}
As a use case for our process (introduced in \cref{fig:eis-for-mm}), we focus on neural networks trained on the MNIST dataset, which consists of 70,000 28x28 grayscale images of handwritten digits from 0 to 9~\cite{726791}. We selected the MNIST dataset because it has become a benchmark for classification tasks in machine learning and computer vision applications~\cite{7966217}. Neural networks were chosen due to their inherent lack of interpretability and their modularity, which allows them to be easily customized by adding layers, changing activation functions, or adjusting other parameters. This flexibility simplifies the process of creating and comparing varied models. Additionally, we apply our novel method for visualizing and, more importantly, comparing multiple models using Chernoff Bots corresponding to these neural networks visualized in a bar chart as introduced in \cref{sec:dashboard}.

\subsection{Preparation: Training a Multitude of Models}
\label{sec:usecase:preparation}
% Paragraaf over hoe wij de training en bijbehorende data collection nu concreet hebben gedaan (zie thesis Gilles?) Aanhalen dat dit het blauwe deel van de afbeelding is, anders is het misschien moeilijk te volgen en dat we ons focussen op het framework dat we voorstellen

% Korte paragraaf die inleid waarom dat we dit doen en hoe
The first stage of the model multiplicity framework is the preparation stage, during which many similar models are trained differently for the same purpose. This approach aims to identify a subset of neural networks that achieve similar accuracy levels but arrive at their decisions through different internal mechanisms, thus belonging to the same Rashomon set. We introduce variations across several dimensions, including model hyperparameters and training data, to induce these differences in the models' internal logic. To facilitate this process, we employ a method that automatically generates models based on random combinations of these variations. \cref{tab:model_variations} outlines the possible values for the variations applied to the training data, model parameters, and training process.

As we want to create models with random variation values, we want to ensure complete coverage of cases or specific scenarios. We partitioned all parameters into equivalence classes. For example, the percentage of typicals is divided into six classes: the lower boundary (0), low (0.2), lower middle (0.4), upper middle (0.6), high (0.8), and the upper boundary (1). Some combinations of these partitions may cause issues, such as when outliers and typicals are set to zero, resulting in no data for training. To address this, we apply constraints, and if a combination is deemed invalid, a new set of values is generated. The following paragraphs discuss and motivate the different variations in more detail.

\begin{table}[ht]
    \centering
    \caption{Overview of the different variations applied to the training data, model parameters, and training process. The table is divided into two sections that summarize the dataset dependent and independent variations within each category and provide an interpretation of their possible values.}
    \label{tab:model_variations}
    \begin{tabularx}{\textwidth}{lX}
        % \hline
        \textbf{Variation}          & \textbf{Possible Values} \\
        \hline
        \multicolumn{2}{c}{\textit{Dataset Independent}} \\
        \hline
        Outlier Percentage          & [0, 0.2, 0.4, 0.6, 0.8, 1]\\
        Typical Percentage          & [0, 0.2, 0.4, 0.6, 0.8, 1] \\
        Number of Hidden Layers     & [1, 2, 3] \\
        Dropout                     & [true, false] \\
        Activation Functions        & [elu, exponential, gelu, hard sigmoid, linear, relu, sigmoid, softmax, swish, tanh] \\
        Batch Size                  & [32, 64, 128, 256, 512] \\
        Optimizer                   & [adadelta, adagrad, adam, adamax, ftrl, nadam, rmsprop, sgd] \\
        Validation Set Usage        & [true, false] \\
        \hline
        \multicolumn{2}{c}{\textit{Dataset Dependent}} \\
        \hline
        Horizontal Translation      & [$-2$, $-1$, 0, 1, 2] \\
        Vertical Translation        & [$-2$, $-1$, 0, 1, 2] \\
        Rotation                    & [\dgr{-20}, \dgr{-15}, \dgr{-10}, \dgr{-5}, \dgr{0}, \dgr{5}, \dgr{10}, \dgr{15}, \dgr{20}] \\
        Contrast Manipulation       & [0, 0.2, 0.4, 0.6, 0.8, 1, 1.2, 1.4, 1.6, 1.8] \\ 
        Image Color Inversion       & [0, 0.2, 0.4, 0.6, 0.8, 1]\\
        \hline
        % \hline
    \end{tabularx}
\end{table}

\paragraph{\textbf{Training dataset variations}}
% Welke variaties?
Since we work with images, we apply data augmentation to enhance the training data. We primarily use position augmentations (translations and rotations) and color augmentations (contrast manipulation and image color inversion) to create variations in the dataset. Translations are limited to a maximum of two pixels in any direction to ensure digits remain within the frame, and rotations are restricted between \dgr{-20} and \dgr{20} in \dgr{5} increments to avoid misinterpreting numbers. This variation enables some models to train with mostly original images, while others are exposed to more inverted images, enhancing their ability to recognize digits across different color schemes.

Image color inversion is achieved by subtracting each pixel's value from 255, effectively transforming light colors to dark and vice versa. A contrast factor of one maintains the original contrast, while values below one reduce the contrast (with a minimum threshold of 0.2 to prevent the image from becoming uniformly black). Values above one increase the contrast up to a maximum of 1.8 to avoid excessive whitening that could obscure important image features. This variation allows some models to train with primarily increased or decreased contrast, leading to different model behaviors. The proportion of these variations in the dataset varies from 0\% to 100\% in increments of 20\%.

% Outlier detection
Additionally, we explore the effects of varying the proportion of outliers and typicals in the training data, ranging from 0\% to 100\%. This approach allows us to observe how models perform under different conditions: some models are trained exclusively with outliers, some with no outliers, and others with a mix of typical data points and outliers. As a result, we obtained a diverse set of models with varying performance, with some excelling in handling outliers and others performing better with typical data. Outlier detection was performed using isolation forests for each digit label~\cite{liu2008isolation}. We chose to identify anomalies within each group of digits rather than across the entire dataset, ensuring that the detection process remains sensitive to the unique characteristics of each digit class and avoids masking subtle outliers specific to individual groups.

\paragraph{\textbf{Model parameter variations}}
The model parameters we selected include the number of hidden layers, the dropout rate, and the activation function. We use between one and three hidden layers, each with 128 neurons, to keep the training duration manageable. Introducing dropout layers helps create model multiplicity by inducing variations in neuron participation, which effectively generates distinct models during each training iteration, even if all other parameters remain the same. For activation functions, we included a range of linear, non-linear, and specialized functions to examine different learning dynamics and performance levels. This variety is particularly relevant for studying model multiplicity, as diverse activation functions can lead to different model behaviors and generalization abilities. It is important to note that only the hidden layers employ variable activation functions, while the output layer consistently uses the softmax function to ensure comparability in predictions. Additionally, dropout and activation functions are consistent across all hidden layers to constrain the variation.

\paragraph{\textbf{Training variations}}
Training variations in our experiment include batch size, optimizer choice, and the use of a validation set. Mini-batch gradient descent is chosen for its advantages in accelerating training and allowing flexibility in batch size adjustment. According to the literature, the optimal batch size typically ranges between 32 and 512~\cite{bengio2012practical,keskar2017on}, so we allowed models to train with batch sizes of 32, 64, 128, 256, and 512 to cover a broad spectrum and enhance model diversity. We considered four categories of optimizers: basic gradient descent variants, adaptive learning rate optimizers, adaptive moment estimation optimizers, and hybrid optimizers. Additionally, we allowed models to either use a validation set or train without one. While a validation set is typically used to assess model performance on unseen data, its use reserves a portion of the training data for validation, which could reduce the amount of data available for training. Given the importance of training data volume in the performance of machine learning models~\cite{uddin2024dataset}, we also opted to include models trained on the full dataset. We set a hard limit of 100 epochs to manage training time across numerous models. Training stops when this limit is reached, even if the model's validation loss continues to improve. This approach ensures models are adequately trained while keeping the overall process more time-efficient.

% Deze sectie eindigt nu wel heel abrupt voor de discussie, eventueel nog iets zeggen over die models (zijn er 500, de accuracy in context van de Rashomon set is xy% etc.) en/of de figuren die eruit kwamen? Eventueel fig 4 en het bijhorende voorbeeld hier zetten in plaats van waar die nu staat?

\begin{figure}[!h]
    \centering
    \includegraphics[width=\textwidth]{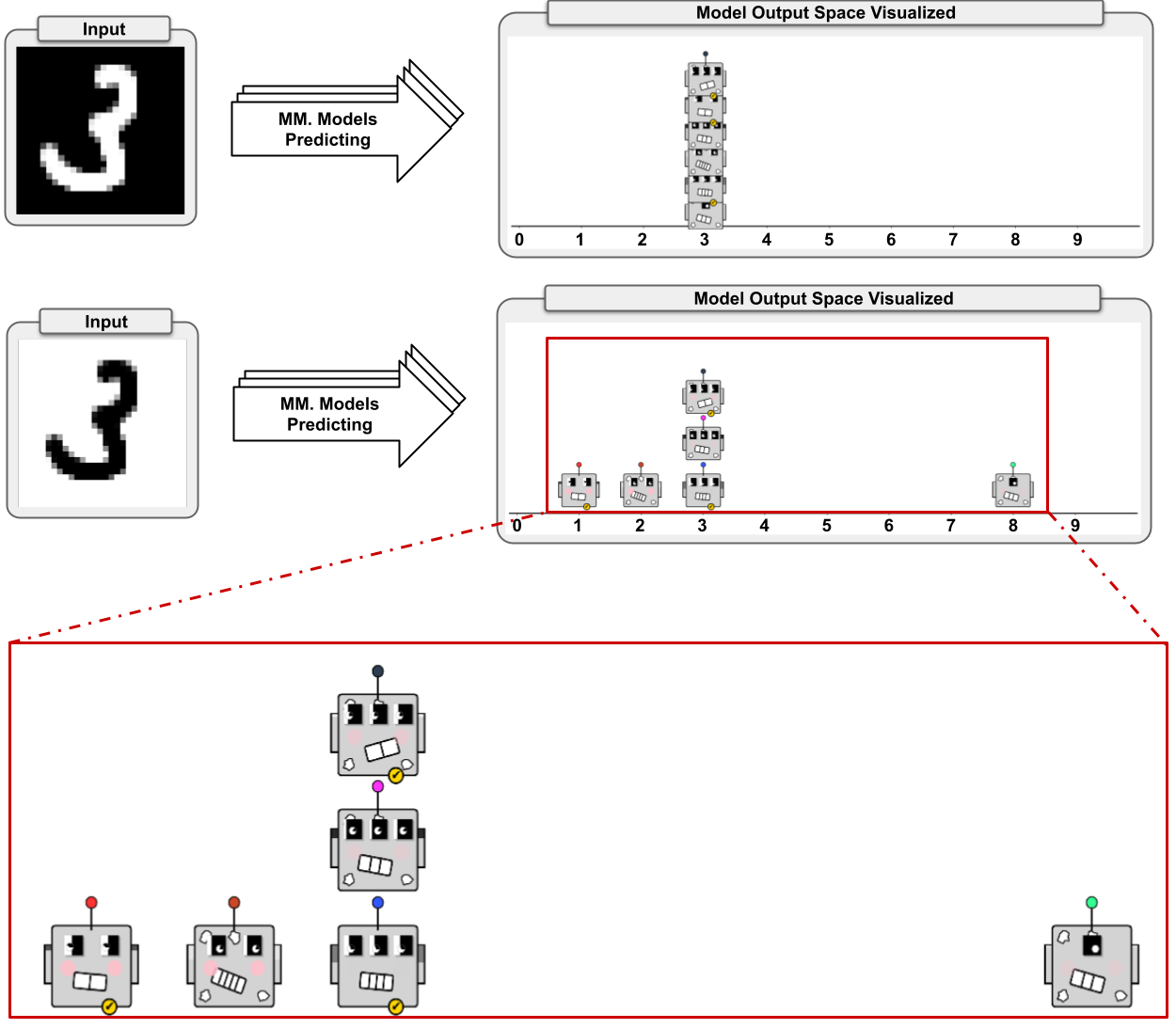}
    \caption{Visualization of model multiplicity predictions on a regular and inverted MNIST sample. The top section shows multiple models predicting the digit ``3'' from a regular MNIST image, where they all agree on the prediction. The middle section illustrates the same models predicting an inverted version of the same digit ``3'', where there is significant disagreement among the models, with predictions spread across different labels. The bottom section zooms in on the inconsistent model predictions, highlighting how model multiplicity results in divergent outputs.}
    \label{fig:chernoffbots_applied}
\end{figure}
As an example of our dashboard, we examine the predictions of a regular and inverted sample of the MNIST dataset in \cref{fig:chernoffbots_applied}. One might initially be inclined to follow the advice of the model that was trained on the most inverted data, which predicts the input label `1' but with low confidence (indicated by prominent red cheeks). Consequently, one might consider more confident models, such as the group of models predicting label `3' or the model predicting `8'. However, the latter model has been trained on barely any inverted data, making it less of an expert in handling inverted data compared to the models that have seen more inverted data during training.

\section{Limitations and Future Opportunities}
In our current solution, we omitted guideline \emph{(G3) ``balance the computational load of multiple models while preserving a positive user experience''} since this is out of scope for this particular work. However, the scalability trade-off between the number of models employed, their respective size and complexity and the importance of the tasks these models support needs to be explored. Increasing the number of models by an order of magnitude must not necessarily correlate with a proportional increase in computational power requirements. For instance, if (a subset of) models are allowed to be less complex as the number of models increases, the overall computational burden could be partially mitigated. For a sustainable and responsible usage of these models, we need to define metrics (e.g., Pareto optimality \cite[Section~3.3]{shoham2008multiagent}) to map the relationship between using more models and the increase in accuracy and trustworthiness.

Our work is focused on black box models, namely classifiers based on neural networks, and the visual dashboard displays the metadata from the models rather than exposing (parts of) the internal behavior of the model. One of the next steps is allowing for a deeper integration of multiple black-box and white-box models in \codename. Since white box models are more interpretable and can provide more detailed explanations, the transparency, auditability and fairness of the overarching system's decision-making will increase by combining these with our current approach. This will require careful balancing of the additional information that is provided to the user with the actual need for more profound insights into model behavior. Furthermore, when the AI models become more context-aware, the need to design solutions that apply guideline \emph{(G16) ``give users insights into how their actions influence each model over time''} will become more important.

Integrating model multiplicity into existing interactive systems necessitates re-evaluating user interface (UI) design; existing UI paradigms must be adapted to effectively communicate the insights of multiple expert models and mechanisms must be developed to aggregate predictions behind the scenes. However, such aggregations may compromise interpretability for end users, raising concerns about transparency in decision-making processes. Consequently, these modifications to established systems would be a costly endeavor. Deploying multiple models simultaneously can lead to inefficiencies in system performance, including increased computational costs and complexities in system maintenance and updates, but also the system's carbon footprint. Therefore, while model multiplicity brings substantial benefits to the robustness and trustworthiness of AI, it also introduces significant engineering challenges that must be addressed to ensure a successful and sustainable implementation and deployment of such systems. Domain experts must carefully \textbf{assess and compare the risk versus the value} of integrating model multiplicity into certain domains.

\section{Conclusion}
% Model multiplicity presents unique challenges in the integration of AI into interactive systems. To address these issues, mainly related to trust and transparency, we have proposed a shift toward a ``many simultaneous expert advisors'' approach. This approach can enhance the accuracy and reliability of AI systems compared to using a single model. Current approaches mainly offer a way to explore the outputs of multiple (typically white-box and tree-based) models in a manner accessible to end users. There are still numerous avenues for further research and development in this field. Future work could focus on expanding the exploration of model multiplicity toward black-box models and addressing the challenges regarding increased computational costs and limited responsiveness to user interaction. To facilitate this process, we designed an initial framework that uses a model multiplicity approach for explainable software engineering.\bigskip

% Contributions:
% (1) We define a process for preparing, training and identifying suitable sets of
% AI models for effective model multiplicity.
% (2) We design a compact and easy-to-recognize visual representation that conveys
% the hyperparameters, properties and background of an AI model to the user.
% For this purpose, we leverage the Chernoff faces technique [11, 18] and tailor
% it to AI models.
% (3) We present a visual dashboard that can present the distribution of answers
% across multiple models in relationship to the varying backgrounds and prop-
% erties of these models, like the overview depicted in Figure 3.

% Kadering + related work
Model multiplicity presents unique challenges in integrating AI into interactive systems, particularly concerning trust and transparency. To address these issues, we proposed \codename that supports a shift toward a ``many simultaneous expert advisors'' approach. This approach enhances the accuracy and reliability of AI systems by leveraging multiple models instead of relying on a single model. %which is essential in critical domains.C%urrent methods often focus on exploring the outputs of various models (typically white-box and tree-based) in a manner accessible to end users. In contrast, our work introduces a process and visual dashboard that supports using and comparing multiple black box AI models and their predictions.

% Contribution 1: process
To address the challenge of integrating model multiplicity into interactive systems, \codename includes a process and visual dashboard that facilitates informed and transparent decision-making. The process is designed to prepare, train and identify sets of multiplicitous models for a given dataset. Specifically, it provides a way to train models with various configurations and demonstrates how these models can be used together within an interactive system \textbf{to counteract the non-deterministic nature of AI}.

% Contribution 2+3: visualization
To effectively present these models to enhance informed, transparent decision-making, we designed a compact and easy-to-recognize visual dashboard by integrating Chernoff Bots, our customized and expandable approach to Chernoff faces, into a bar chart, following the Guidelines for Human-AI Interaction. This visualization approach allows users to quickly understand the hyperparameters, properties and background of an AI model. This transparency \textbf{ensures that humans remain in the loop} so that they can endorse or disregard specific models based on informed judgment selectively. We demonstrate \codename on multiplicitous neural networks for digit classification trained on the MNIST dataset.

Finally, we discussed the limitations of our work and identified future research opportunities, particularly regarding the scalability of our approach to more complex use cases. We believe that investigating the trade-offs between the number of models used, their size and complexity, and the significance of the tasks they address will be crucial for responsibly scaling the use of multiple models. Additionally, we plan to integrate white-box models alongside black-box models in \codename, as we believe this integration would improve the system's overall interpretability. Lastly, we aim to apply \codename within established interactive systems to enhance the transparency of AI systems and facilitate the decision-making of end users. In summary, \codename contributes to the ongoing discourse on enhancing AI systems by promoting greater transparency, trustworthiness, and effectiveness, through the use of multiplicitous models.

\iffalse
To answer the questions identified during the workshop.
1. How can we integrate different engineering approaches that result in one combined architecture or system but have different basic processes?

We propose a process to facilitate the integration and operationalization of multiple model systems within user-interactive systems. This process enables the harmonization of diverse engineering approaches, ensuring that they contribute cohesively to a unified architecture.

% 2. How can we deal with the non-deterministic aspects of AI that have strong implications on usability/utility//reliability?

% By employing a model multiplicitous system, we increase reliability compared to single-model architectures. This approach provides users with multiple outputs from which they can select, thus offering flexibility and a mechanism to counteract the non-deterministic nature of AI.

% 3.Can we ensure that humans remain in the loop?

% We design a system that furnishes the user with comprehensive information to assess the outputs of various models. 

% 4. Can we guarantee the presence of positive hedonic aspects for human users?

% 5. How can we assess the risk versus the value for specific domains?
% The value of working with multiplicitous models is that they provide more holistic insights into possible solutions compared to single-model approaches, which is especially useful in critical domains that risk deciding without this additional knowledge. 
\fi

\begin{credits}
\paragraph{\textbf\ackname}
This work was funded by the Flemish Government under the ``Onderzoeksprogramma Artificiële Intelligentie (AI) Vlaanderen'' programme, R-13509, and by the Special Research Fund (BOF) of Hasselt University, BOF23OWB31.

\paragraph{\textbf\discintname} The authors have no competing interests to declare that are relevant to the content of this article.
\end{credits}

%%
%% The next two lines define the bibliography style to be used, and
%% the bibliography file.
\bibliographystyle{splncs04}
\bibliography{AI-for-ISengineering}

\end{document}